\documentstyle[12pt,epsf,epsfig]{article} 
\def\gsim{\mathrel{\rlap {\raise.5ex\hbox{$ > $}}
{\lower.5ex\hbox{$\sim$}}}}
\def\lsim{\mathrel{\rlap {\raise.5ex\hbox{$ < $}}
{\lower.5ex\hbox{$\sim$}}}}

\newcommand{\mod}{{\bb M}} 
\newcommand{\mods}{{\bbs M}} 

\newcommand{\semiplus}{{\supset\!\!\!\!\!\!\!+~}} 
\newcommand{\be}{\begin{equation}}
\newcommand{\ee}{\end{equation}}
\newcommand{\bea}{\begin{eqnarray}}
\newcommand{\nn}{\nonumber}
\newcommand{\eea}{\end{eqnarray}}
\newcommand{\nd}[1]{/\hspace{-0.6em} #1}

\def\gappeq{\mathrel{\rlap {\raise.5ex\hbox{$>$}}
{\lower.5ex\hbox{$\sim$}}}}
 
\def\lappeq{\mathrel{\rlap{\raise.5ex\hbox{$<$}}
{\lower.5ex\hbox{$\sim$}}}}

\font\mybb=msbm10 at 12pt
\def\bb#1{\hbox{\mybb#1}}
\font\mybbs=msbm10 at 9pt
\def\bbs#1{\hbox{\mybbs#1}}

\begin{document} 
\begin{titlepage} 

\begin{flushright} 
CERN-TH/2000-212 \\
ACT-11/00 \\
CTP-TAMU-23/00 \\
\end{flushright} 

\vspace{0.1in} 
\begin{centering} 

{\Large {\bf  How Large are Dissipative
Effects in Non-Critical Liouville String Theory?}}\\
\vspace{0.2in} 

{\bf John Ellis}$^a$, {\bf N.E. Mavromatos}$^b$ and 
{\bf D.V. Nanopoulos}$^{c}$

\vspace{0.4in} 
{\bf Abstract}

\end{centering} 

{\small 

In the context of non-critical Liouville strings, we clarify why
we expect non-quantum-mechanical dissipative effects to be ${\cal
O}\left(E^2/M_P\right)$, where $E$ is a typical energy scale of the probe,
and $M_P$ is the Planck scale.  In Liouville strings, energy is conserved
{\it at best} only as a statistical average, as distinct from Lindblad
systems, where it is {\it strictly} conserved at an operator level, and
the magnitude of dissipative effects could only be much smaller. We also
emphasize the importance of nonlinear terms in the evolution equation for
the density matrix, which are important for any analysis of complete
positivity.}

\vspace{1.5in}
\begin{flushleft} 

$^a$ Theoretical Physics Division, CERN, CH-1211 Geneva 23, Switzerland
\\
$^b$ Theoretical Physics Group, Department of Physics, 
King's College London, Strand, London WC2R 2LS, U.K. \\
$^c$ Department of Physics, Texas A \& M University, 
College Station, TX~77843, USA; \\
Astroparticle Physics Group, Houston
Advanced Research Center (HARC), 
Mitchell Campus,
Woodlands, TX~77381, USA; \\
Chair of Theoretical Physics, 
Academy of Athens, 
Division of Natural Sciences, 
28~Panepistimiou Avenue, 
Athens 10679, Greece. 
\end{flushleft}

\end{titlepage}

\section{Introduction}

Motivated by speculations concerning quantum gravity,
some years ago a modification of conventional quantum
mechanics was proposed~\cite{ehns}, which admits dissipative
phenomena such as transitions from pure to mixed states.
This approach postulates the appearance of a
non-Hamiltonian term in the quantum Liouville equation
describing the evolution of the density matrix:
\begin{equation}
\dot \rho (t) = i\left[ \rho, H \right] + \nd{\delta H} \rho.
\label{ehns}
\end{equation}
A derivation of this equation based on a
string-inspired treatment of quantum fluctuations in
space-time has also been presented~\cite{emn}.
This formalism was applied initially to simple
two-state systems such as the neutral kaon
system~\cite{ehns,elmn,huet}.
Arguments have been presented that
the magnitude of such dissipative effects could
be suppressed {\it minimally} by a single power of the gravitational
scale $M_P \sim 10^{19}$ GeV, i.e., 
\begin{equation}
\nd{\delta H }={\cal O}\left(\frac{E^2}{M_P}\right)
\label{scale} 
\end{equation}
where $E$ is a typical energy scale
of the probe system which experiences quantum-gravitational 
induced decoherence during its propagation through the 
quantum-gravity `medium'~\cite{emnc}. In particular, the estimate
(\ref{scale}) was argued to apply within a non-critical
Liouville string model for this foamy space-time medium.
If the estimate (\ref{scale}) is indeed correct,
such modifications of quantum mechanics might
be accessible to experiment in the foreseeable future,
for example in the neutral kaon system, which offers one of the most
sensitive microscopic tests of quantum mechanics~\cite{ehns,elmn,huet}.

A new arena for testing quantum mechanics has now
been opened up by neutrino oscillations, and atmospheric-neutrino data
have recently been used~\cite{lisi} to constrain dissipative effects
within a Lindblad formalism~\cite{lindblad}. However,
it has been pointed out~\cite{adler} that the 
order-of-magnitude estimate (\ref{scale}) is 
not applicable to modifications of quantum mechanics
described by the Lindblad formalism~\cite{lindblad},
because the energy scale $E$ appearing in (\ref{scale}) 
is no longer the absolute energy of the probe, but rather {\it the energy
variance} $\Delta E \equiv |E_1-E_2|$ of the two-state system:
\begin{equation}
\nd{\delta H }={\cal O}\left(\frac{\Delta E^2}{M_P}\right)
\label{scale2} 
\end{equation} 
It is therefore crucial~\cite{lisi,adler} to know whether
the Lindblad formalism~\cite{lindblad} applies to the type of
of open quantum-mechanical system provided by a probe
propagating through space-time foam.

There are three key requirements for deriving the Lindblad
formalism~\cite{lindblad}, namely energy conservation at the operator
level, unitarity and entropy production. However, we have pointed out
previously~\cite{emn} within our stringy approach to quantum gravity that,
although unitarity and entropy increase follow from generic properties of
the world-sheet renormalization-group for the Liouville string, energy
conservation should be interpreted {\it at best} as a {\it statistical
property of expectation values}, and even this may be violated in cases
with potential physical interest, such as a $D$-brane model for space-time
foam. Thus we do not believe that the Lindblad formalism~\cite{lindblad}
is directly applicable to propagation through space-time foam, since
energy conservation is not imposed at the operator level. It is this
difference that permits the magnitude of the dissipative effects to be of
order given in (\ref{scale}), larger than that (\ref{scale2})  suggested
in \cite{adler}. We also argue that (\ref{ehns}) contains important
nonlinear effects, which are potentially significant for the
analysis~\cite{benatti} of complete positivity.

\section{Dissipative Dynamics in the Lindblad Formalism}

We first give various definitions 
of concepts used in the following. 
We consider a quantum-mechanical system with a density matrix $\rho$,
in dissipative interaction with an environment.
In the case of a pure state described by a wave vector $|\Psi>$,
the von Neumann density matrix operator is given by:
\be
    \rho = \left| \Psi><\Psi\right|,
\label{densitymatrdef}
\ee
and in a representation $\{ a \}$ the matrix elements of $\rho$ 
are:
\be 
\rho(a,a') \equiv <a'|\Psi><\Psi|a>
\label{repa}
\ee
In general,
for open systems one cannot always define a state vector, in which case 
the density matrix is defined over an ensemble of theories ${\cal M}$:
\be
\rho \equiv {\rm Tr}_{{\cal M}}\left|\Psi><\Psi \right|
\label{opendensity}
\ee
In the coordinate representation: $\{ a \} =\{ \vec x \}$, 
the diagonal element $\rho(x,x;t)$ 
is the probability 
density ${\cal P}(x,t)$, which is given
in the pure-state case by the wave function
at the time $t$:
\be
   \rho(x,x;t) \equiv  {\cal P}(x,t) \left(= \left| \Psi (x,t) \right|^2
\right) \ge 0 
\label{prob}
\ee
For well-defined representations $\{ a \}$
one must have the positivity property
\be 
   \rho(a,a) \equiv \rho_a > 0
\label{positive2}
\ee
Therefore, to define an appropriate density matrix, the operator $\rho_a$
must have positive eigenvalues in the state space $\{ a \}$ at any time
$t$. Even if the matrix $\rho (t=0)$ has positive eigenvalues, it is not
guaranteed that the time-evolved matrix $\rho (t) = \omega (t) \rho(0) $
necessarily also has positive eigenvalues. This requirement of {\it simple
positivity} (SP) has to be imposed, and restricts the general form of the
environmental entanglement. Further restrictions apply when one considers
$\{ a \}$ as coordinates of a subsystem within a larger system interacting
with the environment, namely the requirements of {\it complete positivity}
(CP)~\cite{benatti}.

In the particular case of a linear dissipative system with energy
conservation, the Lindblad
formalism~\cite{lindblad} applies, i.e., the quantum evolution is governed
by a Markov-type process described by an equation of the form:
\bea
&~& \dot \rho (t) = i\left[ \rho, H \right] + \nd{\delta H}[\rho]~, 
\nonumber \\
&~& \nd{\delta H} \equiv 
2\sum_{i}{\cal D}_i^{(d)} 
b_i \rho (t) b_i^\dagger  -
\sum_{i} {\cal D}_i^{(d)}\left( b_i^\dagger b_i \rho (t) + \rho b_i^\dagger b_i \right) ~, \qquad {\cal D}_i^{(d)} \ge 0
\label{lindblad}
\eea
which maintains SP. Here $H$ denotes 
the appropriate Hamiltonian operator of the subsystem,
the positive coefficients ${\cal D}^{(d)}$ are associated to diffusion,
and it can be shown that the extra (non-Hamiltonian) terms 
in the evolution equation (\ref{lindblad}) induce complete decoherence,
as recalled by the index $(d)$~\footnote{It 
should be noticed that the density-matrix evolution equation
(\ref{lindblad}) can be recast, using (\ref{opendensity}),  
as a state-vector evolution equation of stochastic Ito type~\cite{gisin},
if one wishes. We prefer the more general density-matrix formalism,
because the concept of a state vector is
not always well-defined, particularly in our
quantum-gravity foam context~\cite{ehns,emn}.}.

Now suppose one demands that energy be conserved {\it at 
the Hamiltonian operator level}, so that one not only
requires the statistical average 
${\rm Tr}\left(\rho H\right)$ to be independent of time,
but also the absence of explicit time dependence
of the operator $H$:
\be
\frac{\partial H}{\partial t}=0=\frac{d}{dt}{\rm Tr}\left(\rho H\right).
\label{operatorlevelec}
\ee 
This requirement,
together with the monotonic increase of the von Neumann entropy 
$S=-{\rm Tr}\rho{\rm ln}\rho$, implies for the environmental 
operators $b_i,b^\dagger_i$: 
\begin{equation}
\left[ b_i, H \right]=0~, \qquad b_i = b_i^\dagger  
\label{commutdh}
\end{equation}
Then, the environmental part $\nd{\delta H}[\rho]$ 
of the evolution (\ref{lindblad})
assumes a double-commutator form~\cite{zurek,adler}: 
\begin{equation}
\nd{\delta H}=\sum_{i}{\cal D}_i^{(d)}\left[b_i, \left[b_i, \rho \right]\right]
\label{doublecom}
\end{equation}
Clearly, for a two-state system, with energy levels $E_n, n=1,2$,
the only non-trivial Lindblad operators
satisfying (\ref{commutdh}) are of the form $b_i \propto H$.

This simplified case with only one operator
suffices for our purposes~\cite{adler}, and
we restrict our discussion to this case. 
One may then estimate the magnitude of the dissipative effects by 
considering a statistical average of $\nd{\delta H}$ with respect
to a complete basis of states , which can be taken as the
energy eigenstates $\{ |m>\}$ of the Hamiltonian $H$,
$H|m>=E_m|m>$, $m=1,2$. In such a case one has~\cite{adler}:
\begin{equation}
<<\nd{\delta H}>>=\sum_{m,n=1}^{2} <m|\rho|n><n|\rho|m>E_n(E_n-E_m)
=<1|\rho|2><2|\rho|1>(E_2-E_1)^2
\label{variance}
\end{equation} 
and thus one gets the order of magnitude (\ref{scale2}) for
the possible dissipative effects~\cite{adler}.
A similar conclusion is reached if one uses position eigenstates $|q>$ 
as a basis, which is the case of spontaneous localization 
models~\cite{gisin,zurek,adler}, instead of energy eigenstates.
In this case, the dissipative
effects are proportional to the position variance $\Delta q$,
i.e., the separation $|q_1-q_2|$ between the centers of the wave
packets~\footnote{For instance, of the corresponding
neutrino probes in the example discussed in \cite{lisi}.}.
These estimates are considerably smaller than our estimate (\ref{scale}),
so it is important to address carefully the question how
the Liouville formalism differs from the generic Lindblad 
formalism.

\section{Dissipative Dynamics in Non-Critical Strings}

For the benefit of the non-expert reader, we now review how a dissipative
evolution equation of the form (\ref{ehns}) arises in the context of
non-critical string, and how energy is {\it not} conserved at {\it an
operator level}, but {\it at best} as a {\it statistical average}. We also
discuss the special circumstances under which (\ref{lindblad}) may be
obtained, in the hope of clarifying the essential differences between the
two approaches.

\subsection{Non-Critical Strings in Flat World Sheets} 

Critical string theory is described by a conformal field theory $S^*$ on
the two-dimensional world sheet $\Sigma$.
We describe non-critical string in terms of
a generic non-conformal field theory ($\sigma$ model) on 
$\Sigma$. This is perturbed away from conformal
symmetry (criticality) by deformations that are
slightly relevant (in a world-sheet renormalization-group sense),
with vertex operators
$\{ V_i \}$ and couplings $\{ g_i \}$, that parametrize the
space of possible theories:
\be
S=S^* + \int _{\Sigma} d^2\sigma g^iV_i.
\label{l1}
\ee
The renormalization-group $\beta$ function for a coupling $g^i$ on a
flat world sheet reads:
\be
  {\hat \beta}^i = -\varepsilon g^i + \beta^i, \qquad 
\beta^i = C^i_{jk}g^jg^k + \dots 
\label{beta}
\ee
where $\varepsilon \rightarrow 0^+$ is a regularizing parameter,
e.g., in dimensional regularization, the dimensionality
of the world sheet is assumed to be $d=2-\varepsilon$.
As is clear from (\ref{beta}), $\varepsilon$ plays the r\^ole of 
a small anomalous dimension that makes the operator $V_i$
slightly relevant. 
According to standard renormalization theory, counterterms 
can be expanded in poles in $\varepsilon$, and we make the standard
dimensional-regularization assumption
that only single poles matter, whilst higher-order poles cancel
among themselves. 

The fact that the world sheet $\Sigma$ is in general curved
implies that one has to choose a regularization scale $\mu$ 
which is local on the world sheet~\cite{emn}, as is standard 
in stringy $\sigma$ models~\cite{osborn}.
The crucial next step in our approach~\cite{emn} is to promote  
the scale $\mu (\sigma)$ to a dynamical world-sheet field
$\phi$ which should appear in a world-sheet path integral, 
taking the form of Liouville string theory~\cite{liouville}.
Due to its target-space signature, which is {\it negative} for 
the supercritical strings~\cite{aben} we consider,
the world-sheet zero mode of this field is identified
with target time. 

The conformal invariance of the stringy $\sigma$ model
may be restored by Liouville dressing~\cite{liouville},
leading to the following 
equation for the dressed couplings $\lambda^i (\phi,g^i)$ :
\be
  {\ddot \lambda}^i + Q~{\dot \lambda}^i = -{\hat \beta}^i 
\label{lieq}
\ee
where $Q^2 $ is the central-charge deficit of the non-conformal 
theory~\cite{liouville,aben}~\footnote{We used this approach
in~\cite{emn} to derive a stochastic Fokker-Planck 
equation with diffusion for the corresponding 
probability distribution in the theory space $\{ \lambda^i \}$.}.
When one identifies the
Liouville field equation (\ref{lieq}) with a curved-world-sheet
renormalization-group flow~\cite{emn},
the Liouville-dressed couplings $\lambda^i$ may be identified with
appropriately renormalized $\sigma$-model couplings $g_i$ 
on curved world sheets. 

As was shown in~\cite{emn}, the summation over possible topologies
of the world sheet induces canonical
quantization of the theory-space coordinates $\{ g^i \}$, much as
the couplings in local field theories are quantized in the presence of 
wormhole fluctuations in space-time~\cite{coleman}. 
This canonical quantization follows from
a certain set of Helmholtz conditions in theory space~\cite{emn},
which are obeyed provided 
the $\beta^i$ functions obey the gradient-flow property
\be
    {\hat \beta}^i =G^{ij}\frac{\partial C[g]}{\partial g^i} 
\label{threet}
\ee
where $C[g,t]$ is the Zamolodchikov $C$ function~\cite{zam}, 
a renormalization-group invariant combination of 
averages (with respect to (\ref{l1})) of 
components of the world-sheet stress tensor of the deformed $\sigma$ model, 
$G^{ij}$ is the inverse of the Zamolodchikov metric in
theory space: 
\be
    G_{ij} = < V_i V_j > 
\label{metric} 
\ee
and the notation $< \dots >$ denotes any expectation value
over the partition 
function of the deformed $\sigma$ model (\ref{l1}), summed over 
world-sheet genera. 
The Zamolodchikov $C$ function plays the r\^ole of the off-shell 
effective target-space action in string theory, which allows
the identification;
\be
C    [g] =\int dt (p_i {\dot g}^i - E )
\label{Fef}
\ee
where $E$ is the Hamiltonian operator of the string matter.
In critical string theories, the couplings $g^i$ are exactly 
marginal: $\beta^i =0$, and this formalism has trivial 
content, but this is no longer the case~\cite{emn} when one goes
beyond critical strings. 

Renormalizability of the world-sheet $\sigma$ model implies the scale 
independence of physical quantities, such as the partition 
function or the density matrix $\rho(g_i,p^j,t)$ of a string moving in
the background parametrized by the 
$\{ g^i \}$, viewed as generalized coordinates in
string theory space, with the $p_j$ the canonically-conjugate 
momenta in this 
space, that are associated with the vertex operators $V_j$ in a subtle 
sense~\cite{emn}. Let $t={\rm ln}\mu_0$ be the (zero mode
of) the 
world-sheet renormalization-group scale. 
Since the elements of the density matrix are physical quantities, 
and hence independent of the world-sheet scale,
they must obey~\cite{emn} the following renormalization-group equation
 \be
   \frac{d}{dt}\rho(g_i,p^j,t)=\frac{\partial}{\partial t} \rho 
+ {\dot g}^i \frac{\partial}{\partial g^i} \rho + {\dot p}_i 
\frac{\partial}{\partial_i}\rho = 0
\label{renorm}
\ee
where ${\dot {~}}$ denotes a partial derivative with respect to $t$. The
total $t$ derivative above incorporates $t$ dependence both {\it
explicitly} and {\it implicitly}, through running couplings and
generalized momenta~\footnote{Here we study the general consequences
of (\ref{renorm}), although it is possible that only the
diagonal elements of the density matrix in the coordinate representation,
which may be interpreted as probability densities, are in fact
measurable physical quantities, in which case the
requirement (\ref{renorm}) should be applied only 
to the world-sheet partition function of the stringy 
$\sigma$ model.}.
 
As an example how non-critical string dynamics enters, one may consider
the {\it reduced density matrix} $\rho_s (g_s^i,p_{s,j}, t)$
of an {\it effective string theory} describing lowest-level 
string modes characterized by
a {\it subset} $\{ g_s^i \} \in \{ g^i \}$ of couplings to 
operators $\{ V^i_s \}$ that are not exactly marginal.
These deformations are remnants of mixtures of these
lowest-level modes with higher (Planckian) modes of the string, that are
initially exactly marginal. However, this property is lost when the
higher-level modes are integrated over, so that the remnant $\{ V^i_s
\}$ provide a non-trivial
background `environment' for the observable lowest-level modes to 
propagate through.
This `gravitational environment' of higher-level states is
quantized when one sums over world-sheet genera.  
We have made case studies of such systems, based on 
stringy black holes in two space-time 
dimensions~\cite{witten}~\footnote{In this case 
the environment is provided by discrete
delocalized solitonic states, which mix explicitly with lowest-level
propagating matter states of the two-dimensional string in
marginal deformations of the two-dimensional black hole,
as a reflection of infinite-dimensional $W_\infty$ gauge
symmetries~\cite{emn}.}, and higher-dimensional 
analogues provided by the recoil~\cite{emndfoam} 
of $D$ branes~\cite{dbranes} when struck by a light propagating string
state. At present, we only discuss some generic properties
of this Liouville approach to space-time foam~\cite{emn}.

The non-critical subsystem $\{ g_s\}$ 
acquires non-trivial dynamics through the equation (\ref{renorm}),
since the $\beta^i$ functions are non trivial. 
Summing over genera, taking into account the 
canonical quantization of the theory space mentioned above, 
and {\it identifying the Liouville renormalization-group scale
as the target time variable}~\cite{emn}, 
one arrives at the following evolution equation for the 
reduced density matrix {\it operator} $\rho_s$ of
the observable, propagating, localized, lowest-level modes of the 
effective string theory:
\be
\partial _t \rho_s (g_s,p_s,t) = i\left[ \rho_s, {\cal H} \right] 
+ i {\hat \beta _s}^i G_{ij} \left[g^j_s, \rho_s\right] 
\label{liouvilledens}
\ee
where ${\cal H}$ is a Hamiltonian operator for the subsystem 
consisting of propagating string modes, and the second term on the
right-hand side of (\ref{liouvilledens}) is an explicit
string representation for $\nd {\delta H}$ in (\ref{ehns}). 
It is clear that our fundamental equation (\ref{liouvilledens}) stems from  
the requirement of renormalizability (\ref{renorm}).  

This description has several important properties, which we now describe.

$\bullet$ {\it Probability conservation}:
The total probability $P=\int dp_l dg^l Tr[\rho(g^i, p_j)]$ is
conserved, because
\be
{\dot P}=\int dp_l dg^l Tr[\frac{\partial }{\partial p_i}
(G_{ij}\beta^j \rho )]
\label{fivet}
\ee
can receive contributions only from the boundary of phase space,
that must vanish for an isolated system.

$\bullet$ {\it Entropy growth}:
The entropy $S=-Tr(\rho ln \rho)$ is 
not conserved, to the extent that relevant couplings with
$\beta^i \ne 0$ are present:
\be
{\dot S}=(\beta^i G_{ij} \beta^j)S
\label{sevent}
\ee
implying a monotonic increase for unitary world-sheet theories
for which $G_{ij}$ is positive definite. We see from (\ref{sevent})
that {\it any} running of {\it any} coupling will lead to an
increase in entropy, and we have interpreted \cite{emn} this
behaviour in terms of quantum models of friction. The
increase (\ref{sevent}) in the entropy corresponds to a
loss of quantum coherence, which is also known in
models.

$\bullet$ {\it Statistical conservation of energy}:
The most important property for our purposes
here is that 
energy is {\it at best} conserved statistically {\it on the
average}~\cite{emn}.
In other words, one has explicit time-dependence 
(dissipation) in the Hamiltonian operator of the subsystem,
thereby allowing flow of energy to the environment
at an operator level. Then, it is {\it at best} 
only the right-hand equality in (\ref{operatorlevelec})
that is valid, depending on the specific model~\footnote{See
the next section for a question-mark that hangs over even this 
statistical equality.}. 

To see this, we recall, as discussed above (\ref{renorm}), that
the renormalizability of the $\sigma$ model implies
$d/dt{\rm Tr}\rho =0$. Since we identify the target time with the
renormalization-group scale $t$ on the world sheet~\cite{emn},
$\partial _t {\rm Tr}\rho_s$ may be expressed 
in terms of the renormalized couplings $g^i$ by means of the 
evolution equation (\ref{liouvilledens}). We now compute
\be
\frac{\partial}{\partial t}
<<E>>=
\frac{\partial}{\partial t} Tr(E\rho) = <<\partial _t (E -
\beta^i G_{ij} \beta^j)>>
\label{energcons}
\ee
where $E$ is the Hamiltonian operator, and
$<< \dots >> \equiv Tr[\rho (\dots)]$.
In deriving this result, we took into account 
the evolution equation (\ref{liouvilledens}),  
and the quantization rules in theory space~\cite{emn},
\be
[g^i, g^j]=0 \qquad ; \qquad [g^i, p^j]=-i\delta^{ij}
\label{ccs}
\ee
as well as the fact that in string $\sigma$ models
the quantum operators $\beta^i G_{ij}$ are
functionals only of the coordinates $g^i$,
and not of the generalized momenta $p^i$. 
We also note that, in  our approach,
the total time derivative of an operator ${\hat Q}$ is given as
usual by
\be
\frac{d}{dt} {\hat Q} =-i[ {\hat Q}, E] + \frac{\partial {\hat Q}}{\partial t}
\label{timederivative}
\ee
We recall that total time derivatives
incorporate both explicit and implicit 
renormalization-scale dependence (via running
couplings), whilst
partial time derivatives incorporate only the explicit dependence.

Using the $C$-theorem results  (\ref{threet},\ref{Fef})~\cite{zam,emn}
and the formalism developed in~\cite{emn},
it is straightforward to arrive at
\be
\frac{\partial}{\partial t}
<<E>>=
\frac{\partial}
{\partial t} (p_i\beta^i) 
\label{sixt}
\ee
In conventional stringy $\sigma$ models,  
due to world-sheet renormalizability
with respect to a `flat' world-sheet cut-off, 
any dependence on the renormalization
group scale in the $\beta^i$ functions
is implicit through the renormalized couplings,
and hence $\partial_t \beta^i=0$. Moreover, 
the quantity $p_i$ appearing in (\ref{sixt}) may be written in the form
\be
    p_i =G_{ij}\beta^j \sim \sum_{n} C_{ii_1 \dots i_n}g^{i_1}\dots g^{i_n}
 \label{momenta}
\ee
where the $C_{ij\dots }$ are the (totally symmetric) correlators of 
vertex operators $<V_iV_j \dots >$. In the usual case,
due to the renormalizability of the $\sigma$-model theory,
there is no explicit dependence on the world-sheet scale $t$ 
in such correlators, neither on $\beta^i$, and hence the 
right-hand side of (\ref{sixt}) vanishes, implying 
{\it energy conservation  on the average}.

In this derivation, {\it renormalizability} replaces the
{\it time-translation
invariance} of conventional target-space field theory.

\subsection{Curved World-Sheet Renormalization and
Generic Liouville Correlators}

An additional feature appears
in certain non-critical string theories that involve solitonic 
structures in their backgrounds, such as $D$
particles~\cite{kogan,emndfoam,mavro+szabo}.
There are deformations in the set of $\{ V_i \}$ 
that obey a logarithmic conformal algebra~\cite{lcft},
rather than an ordinary conformal algebra. 
As discussed in detail in~\cite{mscth}, 
the field correlators $C_{i_1 \dots i_m}$ in such logarithmic 
conformal field theories
exhibit {\it explicit} dependences on the world-sheet renormalization
(time) scale $t$. This, in fact, is essential in guaranteeing the 
gradient-flow property (\ref{threet}) of the 
corresponding $\beta$ functions~\cite{mscth},
which is crucial for canonical 
quantization in theory space~\cite{emn}, as we discussed previously.
As a result, our previous argument for
energy conservation on the average breaks down in theories with
logarithmic deformations, because
the right-hand-side of (\ref{sixt}) 
is no longer non-zero. Physically, this is explained by the flow of energy 
from the propagating subsystem to the recoiling $D$--particle 
background~\cite{mavro+szabo}.

In standard (critical) string theory, the correlators $C_{i_1 \dots i_m}$ 
are associated with elements in the $S$ matrix for
particle scattering, and as such should not exhibit
any explicit dependence on the time coordinate.
In view of their explicit dependence on the world-sheet scale 
in logarithmic conformal field theories, which in our approach~\cite{emn}
is identified with the target time, such correlators cannot be 
interpreted as conventional scattering amplitudes in target space,
but rather as non-factorizable $\nd{S}$ matrix elements. The time
dependence also means that,
whilst the initial formulations of quantum-gravitational 
dissipation~\cite{ehns,elmn,huet} assumed energy conservation,
this can no longer be guaranteed
in string soliton models of space-time foam~\cite{emndfoam}.

There is one more formal reason for relaxing strict
energy conservation in Liouville strings,
which we review below~\cite{mn}. The discussion of
the previous subsection pertained to flat world-sheets,
but the situation 
is different when one considers generic correlators in
Liouville strings, because the world-sheet 
has curvature expressed essentially by the dynamical Liouville mode. 
A more correct approach is to consider renormalization
in curved space~\cite{osborn}, which leads to 
new types of counterterms. 
It is just this feature that may lead to violations of 
the energy conservation, as happens explicitly in the
$D$--brane case~\cite{mavro+szabo,mscth}, which is a particular
case of Liouville strings. Here we
briefly review the situation,
concentrating on those aspects of the formalism 
relevant to energy conservation, referring the reader
interested in more details to the literature~\cite{mn,emn}. 

\begin{figure}[htb]
\begin{center}
\epsfig{figure=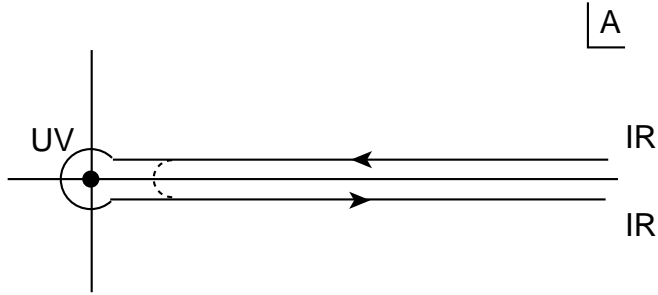}
\caption{\it The contour
of integration for the analytically-continued
(regularized) version of $\Gamma (-s)$ for $ s \in Z^+$.
The quantity $A$ denotes the (complex) world-sheet area. 
This is known in the literature as the Saalschutz contour,
and has been used in
conventional quantum field theory to relate dimensional
regularization to the Bogoliubov-Parasiuk-Hepp-Zimmermann
renormalization method. Upon the interpetation of the 
Liouville field with target time, this curve
resembles closed-time paths in non-equilibrium field theories~\cite{ctp}.} 
\label{fig1}
\end{center}
\end{figure}

We consider the $N$-point correlation function of vertex operators 
in a generic Liouville theory, viewing the Liouville field
as a local renormalization-group scale on the world sheet~\cite{emn}.
Standard computations~\cite{goulian} yield the following form
for an $N$-point correlation
function of vertex operators integrated over the world sheet:
$V_i\equiv \int d^2z V_i (z,{\bar z}) $,
\be
A_N \equiv <V_{i_1} \dots V_{i_N} >_\mu = \Gamma (-s) \mu ^s
<(\int d^2z \sqrt{{\hat \gamma }}e^{\alpha \phi })^s {\tilde
V}_{i_1} \dots {\tilde V}_{i_N} >_{\mu =0}
\label{C12}
\ee
where the tilde denotes removal of the zero mode of the
Liouville  field $\phi $. The world-sheet scale $\mu$ is associated with
cosmological-constant terms on the world sheet, which are characteristic
of the Liouville theory, and the
quantity $s$ is the sum of the Liouville anomalous dimensions
of the operators $V_i$:
\be
s=-\sum _{i=1}^{N} \frac{\alpha _i}{\alpha } - \frac{Q}{\alpha}
\qquad ; \qquad \alpha = -\frac{Q}{2} + \frac{1}{2}\sqrt{Q^2 + 8}
\label{C13}
\ee
The $\Gamma $ function appearing in (\ref{C12}) can be
regularized~\cite{kogan2,emn} for negative-integer
values of its argument by
analytic continuation to the complex plane using the
the Saaschultz contour of Fig.~\ref{fig1}. 

To see technically why the above formalism 
leads to a breakdown in the 
interpretation of the correlator $A_N$ 
as a string amplitude or $S$-matrix element, which in turn leads to  
the interpretation of the world-sheet
partition function as a probability density rather than a 
wave function in target space, 
one first expands the Liouville
field in (normalized) eigenfunctions  $\{ \phi _n \}$
of the Laplacian $\Delta $ on the world sheet
\be
 \phi (z, {\bar z}) = \sum _{n} c_n \phi _n  = c_0 \phi _0
 + \sum _{n \ne 0} \phi _n \qquad \phi _0 \propto A^{-\frac{1}{2}}
\label{C14}
\ee
where $A$ is the world-sheet area, and
\be
   \Delta \phi _n = -\epsilon_n \phi _n  \qquad n=0, 1,2, \dots,
\qquad \epsilon _0 =0
\qquad (\phi _n, \phi _m ) = \delta _{nm}
\label{C15}
\ee
The correlation functions (without the Liouville
zero mode) appearing on the right-hand side of (\ref{C12})
take the form~\cite{mn}
\bea
{\tilde A}_N \propto &\int & \Pi _{n\ne0}dc_n exp(-\frac{1}{8\pi}
\sum _{n\ne 0} \epsilon _n c_n^2 - \frac{Q}{8\pi}
\sum _{n\ne 0} R_n c_n + \nn \\
~&~&\sum _{n\ne 0}\alpha _i \phi _n (z_i) c_n )(\int d^2\xi
\sqrt{{\hat \gamma }}e^{\alpha\sum _{n\ne 0}\phi _n c_n } )^s
\label{C16}
\eea
where $R_n = \int d^2\xi R^{(2)}(\xi )\phi _n $. We can compute
(\ref{C16}) if we analytically continue \cite{goulian}
$s$ to a positive integer $s \rightarrow n \in {\bf Z}^{+} $.
Denoting
\be
f(x,y) \equiv  \sum _{n,m~\ne 0} \frac{\phi _n (x) \phi _m (y)}
{\epsilon _n},
\label{fxy}
\ee
one observes that, as a result
of the lack of the zero mode,
\be
   \Delta f (x,y) = -4\pi \delta ^{(2)} (x,y) - \frac{1}{A}.
\label{C17}
\ee
We may choose
the gauge condition  $\int d^2 \xi \sqrt{{\hat \gamma}}
{\tilde \phi }=0 $, which determines the conformal
properties of the function $f$ as well as its
`renormalized' local limit
\be
   f_R (x,x)=lim_{x\rightarrow y } (f(x,y) + {\rm ln}d^2(x,y))
\label{C18}
\ee
where  $d^2(x,y)$ is the geodesic distance on the world sheet.
Integrating over $c_n$, one obtains
\bea
~&& {\tilde A}_{n + N} \propto
exp[\frac{1}{2} \sum _{i,j} \alpha _i \alpha _j
f(z_i,z_j) + \nn  \\
~&&\frac{Q^2}{128\pi^2}
\int \int  R(x)R(y)f(x,y) - \sum _{i} \frac{Q}{8\pi}
\alpha _i \int \sqrt{{\hat \gamma}} R(x) f(x,z_i) ]
\label{C19}
\eea
We now consider
infinitesimal Weyl shifts of the world-sheet metric,
$\gamma (x,y) \rightarrow \gamma (x,y) ( 1 - \sigma (x, y))$,
with $x,y$ denoting world-sheet coordinates.
Under these Weyl shifts,
the correlator $A_N$
transforms as follows~\cite{mn,emn} 
\bea
&~&
\delta {\tilde A}_N \propto
[\sum _i h_i \sigma (z_i ) + \frac{Q^2}{16 \pi }
\int d^2x \sqrt{{\hat \gamma }} {\hat R} \sigma (x) +    \nn \\
&~&
\frac{1}{{\hat A}} \{
Qs \int d^2x \sqrt{{\hat \gamma }} \sigma (x)
       +
(s)^2 \int d^2x \sqrt{{\hat \gamma }} \sigma (x) {\hat f}_R (x,x)
+  \nn \\
&~&
Qs \int \int d^2x d^2y
\sqrt{{\hat \gamma }} R (x) \sigma (y) {\hat {\cal
 G}} (x,y) -
  s \sum _i \alpha _i
  \int d^2x
  \sqrt{{\hat \gamma }} \sigma (x) {\hat {\cal
 G}} (x, z_i) -   \nn \\
&~&
 \frac{1}{2} s \sum _i \alpha _i{\hat f}_R (z_i, z_i )
  \int d^2x \sqrt{{\hat \gamma }} \sigma (x)
-    \nn \\
&~&
 \frac{Qs}{16\pi} \int
  \int d^2x d^2y \sqrt{{\hat \gamma (x)}{\hat \gamma }(y)}
  {\hat R}(x) {\hat f}_R (x,x) \sigma (y)\} ] {\tilde A }_N
\label{dollar}
\eea
where the hat notation denotes transformed quantities,
and
the function  ${\cal G}$(x,y)
is defined as
\be
  {\cal G}(z,\omega ) \equiv
f(z,\omega ) -\frac{1}{2} (f_R (z,z) + f_R (\omega, \omega ) )
\label{C20}
\ee
and transforms simply under Weyl shifts~\cite{mn,emn}.
We observe from (\ref{dollar}) that
if the sum of the anomalous dimensions
$s \ne 0$, the `off-shell' effect of
non-critical strings, then there are
non-covariant terms in
(\ref{dollar}), inversely proportional to the
finite-size world-sheet area $A$.
Thus the generic correlation 
function $A_N$ does not have a well-defined finite
limit as $A \rightarrow 0$. 

In our approach to string time, we identify~\cite{emn} 
the target time as $t=\phi_0=-{\rm log}A$, 
where $\phi_0$ is the world-sheet zero mode of the Liouville field.
The normalization is specified by requiring
the canonical form of the kinetic term for $\phi$ 
in the Liouville $\sigma$ model~\cite{aben,emn}. 
The opposite flow of the target time, as compared to that of the 
Liouville mode, follows from the `bounce' picture~\cite{kogan2,emn} for
Liouville flow of Fig.~1.
The induced time- (world-sheet scale 
$A$-) dependences
of the correlation functions $A_N$ imply the
breakdown of their interpretations as
well-defined $S$-matrix elements,
whenever there is a departure from criticality: $s \ne 0$. 

We now obtain a non-zero contribution to (\ref{sixt}) from
an apparent non-trivial explicit $t$ dependence in
$\partial_{g^i}\beta^i$
through the coefficients $C^i_{i_1 \dots i_m}$ of the 
$\beta^i$ functions, associated with the breakdown of their
identification
as well-defined $S$-matrix elements, in contrast to the conventional 
string case. 
As discussed in~\cite{emn}, one can explicitly verify this 
picture in the case of $D$--particle recoil.
The scattering of a low-energy (string) matter particle off a
$D$ particle conserves energy {\it only} in the complete system,
when the recoil of the $D$ particle is taken properly into 
account~\cite{mavro+szabo}. The recoil degrees of freedom 
are entangled with the subsystem of low-energy string matter,
and their neglect leads to an explicit violation of the energy
conservation condition at an operator level.
Energy conservation can {\it at best} be imposed 
in a {\it statistical} average sense.

\subsection{Can one Restore the Lindblad Formalism in Non-Critical
String?}

We have seen that the non-critical-string time evolution
(\ref{liouvilledens}) is nothing other than a world-sheet
renormalization-group evolution equation in coupling constant space $\{
g^i \}$ for the $\sigma$ model, and, in view of the failure of energy
conservation, discussed above, cannot in general be written in the
Lindblad form (\ref{lindblad}). There are, however, some special cases in
which the equation can indeed be put in a Lindblad form, but even in such
cases, as we shall explain below, the order of magnitude of the
dissipative effects is still given by (\ref{scale}) and {\it not}
(\ref{scale2}). 

We first recall from our above discussion
that the non-critical string evolution equation
(\ref{liouvilledens}) represents diffusive evolution in theory space of
the non-critical string $\{ g^i \}$. Hence, as explained in detail
in~\cite{emn}, there is spontaneous localization in such a 
space~\cite{gisin,zurek}. 
Thus, even if the situation resembled that of Lindblad, it would 
not have been associated with energy-driven diffusion, as was the case 
discussed in \cite{lisi}, but with spontaneous localization
models~\cite{adler}.  
This can be seen straightforwardly from the 
form (\ref{liouvilledens})
for the time evolution of the density matrix
in non-critical string theory.
In this case, there is no `environment operator' that commutes with the 
Hamiltonian, for the simple reason that the r\^ole 
of $b_i$ is played in Liouville strings by
various partitions (within a generalized definition 
of the quantum-ordering prescription for the operators $g^i$) of 
$G_{ij}\beta^j = 
C_{ii_1 \dots i_m}g^{i_1} \dots g^{i_m}$. The operators
$g^i$ play the r\^ole of position operators in a generalized 
coupling-constant space, and, as such, the $g^i$ do not commute with $H$
in general, which depends on the generalized momenta in theory space $p_i$. 
Hence we does not respect the Lindblad criterion (\ref{commutdh}). 

It can easily be seen that a
double-commutator structure could {\it only} arise in a 
situation where {\it only} the linear anomalous-dimension terms
are kept
in $\beta^i=y^ig^i + \dots$, with no sum over $i$. In that case, 
making the antisymmetric ordering prescription denoted by $: \dots :$,
the diffusive term in the Liouville string evolution 
equation (\ref{liouvilledens}) does acquire a double-commutator
structure 
\be
:\nd{\delta H}:~ \sim~ y^i [g^i, [g^i,\rho]] 
\label{doublecommliouv}
\ee
In the two-state case studied in~\cite{lisi,adler},
this leads to an estimate of the dissipative effects as being of order
${\cal O}\left( y^i (\Delta g^i)^2\right)$,  
where $\Delta g^i$ is the variance in 
theory space, which should be {\it distinguished} from real position
variances. 
Indeed, in real space, $\Delta g^i $ may be considered of order one,
and this leaves the anomalous dimension factor $y^i$ to determine the
order
of the effect. This is due to the fact that, as discussed in~\cite{emn},
the Liouville string case is characterized by the appearance of
pointer states~\cite{zurek} in theory space, namely
wave-packets with $\Delta g^i \sim h_s$, where $h_s$ is the
`Planck constant' in theory space, which is found~\cite{mavro+szabo} 
to be proportional to the string coupling $g_s$.
This latter property can easily be understood
from the fact that quantization in theory space 
is induced in our approach by higher-genus topologies on the 
world-sheet, and hence string-loop interactions~\cite{emn}. 
In conventional string models, $g_s$ is 
of order one, namely $g_s^2/4\pi=1/20$. However, 
one may consider more general cases in which
$g_s$ is a phenomenological parameter, to be constrained by data,
e.g., those on neutrinos~\cite{lisi}. 

In generic non-critical string models,
the operators corresponding to the $g^i$ are $(1,1)$
on the world sheet, so that $y^i=0$, but {\it not exactly marginal},
which implies  that only the three- and higher-point-function
contributions are non zero in 
$\beta^i=C^i_{jk}g^jg^k + \dots $. The latter terms clearly
do not lead to a double commutator structure 
(\ref{doublecommliouv},\ref{doublecom}).
The order of the corresponding effects can, however, be estimated
by the fact that the correlators $C^i_{i_1 \dots i_m}$ 
are viewed as (non-factorizable) 
$\nd{S}$-matrix elements in target space, and as such 
can be expanded in a power series in $\alpha' k^2$ in the
interesting closed-string case, where $k$ is a 
typical target-space four-momentum scale, 
$\alpha'=1/M_s^2$ 
the Regge slope, and $M_s$ the string scale. This yields
once more the estimate (\ref{scale}) for string-induced 
dissipative effects. 

This estimate is supported 
by many specific examples. For instance, 
in the context of the two-dimensional black-hole 
model of~\cite{witten}, the analysis of~\cite{emn} showed that the 
{\it exactly marginal} world-sheet correlators involve
{\it necessarily} the coupling of low-energy
propgating modes with Planckian solitonic modes. The 
world-sheet correlators involving the latter are 
suppressed to leading order
by a single power of the Planck mass $M_P$~\footnote{Strictly speaking, 
the string scale $M_s$, but we do not draw the distinction here.}. 
This coupling is necessitated by stringy gauge symmetries,
specifically $W_\infty$ 
symmetries, which were argued in~\cite{emn} to be 
responsible for maintaining quantum coherence at the microscopic string
level, but not at the level of the low-energy effective
theory relevant to observation. This 
in turn implies that the splitting between low-energy propagating modes,
and quantum-gravitational modes is suppressed by a single inverse power 
of $M_s$ or $M_P$, leading again to the above estimate for 
the magnitude of $C^i_{i_1\dots i_n}$,
where the $g^{i_n}$ refer to low-energy propagating matter modes. 
In a similar spirit, the recoil approach to $D$--brane/string 
scattering~\cite{kogan,emndfoam}, which is another example of 
a gravitational medium, also suggests that dissipative effects
are suppressed by a single inverse power of the gravitational mass scale, 
as in (\ref{scale}).

It goes without saying, however, that there is always the possibility of a
cancellation in some specific case, so that the
suppression is by some higher power of the
gravitational scale as in (\ref{scale2}), but we see so reason why this
should be generic. 

\section{Nonlinearities and Complete Positivity}

We conclude with some remarks on nonlinearities in Liouville string
dynamics.  It has been pointed out~\cite{benatti} that the naive extension
of a single-particle approach to two-particle systems may not respect
complete positivity (CP), and constraints on the parametrization of
dissipative effects in single-particle systems have been proposed.
However, as we argue below, it may be an oversimplification to ignore the
likelhiood of nonlinearities in the quantum-gravitational framework, that
would require the issue of complete positivity~\cite{benatti} to be
re-evaluated.

An important indication of the possible importance of nonlinear
environmental effects comes from the form of the evolution equation
(\ref{liouvilledens}) for the reduced 
density matrix in the context of Liouville strings.
The appearance of the
$G_{ij}\sim <V_iV_j>$ term given by (\ref{metric}) 
in the dissipative part
of (\ref{liouvilledens}) is a signature of
{\it nonlinear} Hartree-Fock evolution, since
the expectation values
$< \dots >$ are to be evaluated with respect to a 
world-sheet partition function $\Psi [g^i, t]$ 
of the string that is resummed over genera. 
According to the detailed discussion in~\cite{msnonlinear}, such 
a resummed world-sheet
partition function may be identified, as the trace of the density matrix
$\rho$, with a probability distribution in theory space. As such,
the dissipative aspects of the evolution exhibit 
a nonlinear integro-differential form:
\be
\partial _t \rho_s \ni 
\left({\rm Tr}_{g_i}\rho_s~V_iV_j \right){\hat \beta}^j [g^j_s, \rho_s]
\label{integrodiff}
\ee
where ${\rm Tr}_{g_i}$ denotes a partial trace over 
quantum fluctuations about 
the string background $\{ g_i \}$~\cite{emn,mavro+szabo}:
\be
{\rm Tr}_{g_i}\left(\dots \right) \equiv \int 
\sqrt{\Gamma^{-1}}d\alpha_i e^{-\frac{\alpha_i G^{ij} 
\alpha_j}{\Gamma}}e^{-S^* + \int _{\Sigma} (g_i + 
\alpha_i)V_i}\left( \dots \right)
\label{beurk}
\ee
and the quantum fluctuations $\alpha_i$ are of stochastic 
type, with a Gaussian probability distribution in theory space,
as a result of the sum over world-sheet topologies. 

The equation (\ref{integrodiff}) should be understood as an 
operator equation in theory space. 
Its form is consistent with the fact that, in the $\{ g_i \}$ 
representation, 
$G_{ij} \sim \frac{\delta^2}{\delta g_i \delta g_j}{\rm Tr}_{g_i}\rho_s$,
as follows from canonical quantization in theory space~\cite{emn,mavro+szabo},
according to which $V_i \rightarrow -i\delta /\delta g^i$. 
Near a fixed point in theory space, as is
reached at asymptotically large 
times $t \rightarrow \infty$,
where perturbation theory in $\{ g_s \}$ is valid, 
it might be a sufficiently good approximation to ignore the
nonlinearities
and parametrize the terms $G_{ij}\beta^j$ by some `constants'. 
However, one should not expect this linearization
to be valid throughout the evolution, e.g., at early times 
after  
the production in $\phi$ decay of a correlated $K^0 - {\bar K}^0$ system.

We do not pursue this point here, but emphasize that, in our view,
such nonlinearities need to be understood before using
complete positivity to impose
restrictions on the parametrization of $\nd{\delta H }$ for
single-particle systems that go
beyond those imposed by simple positivity~\cite{benatti}.
They may also be important for the understanding of the energy-momentum
conservation issue raised in~\cite{huet}.

\section{Outlook on Neutrino and other Experiments}

We have reviewed in this paper why we expect that dissipative
effects on isolated systems due to quantum-gravitational effects
might be as large as ${\cal O}(E^2/M_P)$. This estimate is
potentially very encouraging for several classes of experiments,
possibly including neutrinos. 

As has been discussed extensively,
present and near-future kaon experiments may be sensitive to an effect of
this magnitude~\cite{ehns,elmn,huet}. The DAFNE experiments on
$\phi \rightarrow K^0 + {\bar K}^0$ are particularly interesting in this
regard, because of the two-particle correlation at production.
However, as just mentioned, we need more understanding of the
possible impact of nonlinear effects in this case.

The suggestion~\cite{lisi,Klap} that neutrinos might be sensitive to
dissipative quantum-gravitational effects is very
interesting~\footnote{There is also a suggestion to probe these 
via double-$\beta$ decay experiments~\cite{Klap}.}. As discussed
above, we do not believe that the Lindblad formalism
is necessarily applicable, at least in the form discussed so far in the
literature~\cite{lisi,adler}. A different issue is that of Lorentz
invariance. The formalism of~\cite{ehns,emn} is {\it not} Lorentz
invariant, and we have suggested an approximate treatment for
ultra-relativistic particles such as photons~\cite{amelino} or
neutrinos~\cite{volkov}, which we believe to be the most appropriate
starting-point for an analysis of neutrino data~\cite{emnc}. In this
approach,
the velocity of a photon (or massless) neutrino deviates from $c$,
which is identified as the low-energy limit of the velocity of light:
\begin{equation}
v \, = \, c \times \left[ 1 \, - \, {E \over M} \, + \dots \right]
\label{modvel}
\end{equation}
where $E$ is the energy and $M$ is some large mass scale that might be
${\cal O} (M_P)$.
This deviation from $c$ has the characteristics of a {\it refractive
index} in vacuo. In addition, there may be stochastic fluctuations 
in the velocity of a photon (or neutrino) of specified energy,
that have a {\it diffusive} character. These suggestions arise from our
considerations of recoil effects on quantum-gravitational vacuum
fluctuations due to the passage of an energetic particle~\cite{emnc}.
Timing observations
of distant astrophysical sources are sensitive to $M \sim 10^{15}$~GeV,
and there are prospects to increase this to $M \sim M_P$ or
beyond~\cite{amelino}.

If only the refractive index effect (\ref{modvel}) is present, and the
quantum-gravitational mass parameter $M$ is flavour-independent, there
would be no practical consequences for neutrino oscillation physics.
However, consequences would ensue if $M$ is flavour-dependent, or if
there are also diffusive effects. We plan to return to these issues in a
future publication.

\section*{Acknowledgements} 

We thank Eligio Lisi for arousing our interest in this question.
The work of N.E.M. is partially supported by P.P.A.R.C. (U.K.)
through an Advanced Research Fellowship.
That of D.V.N. is partially supported by DOE grant 
DE-F-G03-95-ER-40917.
N.E.M. and D.V.N. also thank
H. Hofer for his interest and support.

\end{document}